%% file: w51_review.tex
\documentclass{emulateapj}

\input{preface}

\begin{document}
\title{A review of the W51 cloud}
\titlerunning{W51}
\authorrunning{Ginsburg}
\input{authors}

\begin{abstract}
The W51 cloud complex is one of the best laboratories in our Galaxy to study
high-mass star formation.  At a distance of about 5 kpc, it is the closest
region containing a high-mass protocluster, and it has two.  The cloud includes
a long infrared-dark cloud, is interacting with a supernova remnant, and
contains a variety of unique massive protostellar sources.  This article is an
observational review of the region.

\end{abstract}
\ifpdf
\maketitle
\fi


\section{Overview}
The W51 giant molecular cloud is among the most massive and active star-forming
regions in our Galaxy.  While it was originally discovered as a bright radio
source and identified as an \hii region \citep{Westerhout1958a,Mehringer1994a},
it has since become notable as an extremely gas- and dust-rich cloud
\citep{Carpenter1998a,Ginsburg2012a,Urquhart2014a,Wang2015a}.  It is
particularly notable for its two most luminous high-mass protostars, W51e2 and
W51 North, both of which exhibit extreme chemical richness and are sites of
uncommon masers \citep{Zhang1997a,Eisner2002a,Shi2010a,Henkel2013a,Goddi2015a}.
This review will discuss W51 in both a Galactic context and in its role as
a laboratory for high-mass star formation studies.


\section{W51 in the context of the Galaxy}
Our Galaxy contains only a few molecular clouds with $M\gtrsim10^6$ \msun, and
these clouds dominate the molecular mass in the Galaxy \citep{Combes1991a}.  Of
this sample, W51 is perhaps the most observationally isolated, located in a
region of the galaxy with little foreground or background material around
$\ell=49.5, b=-0.4$.  Its location has made it an appealing target for
large-scale surveys in CO
\citep{Carpenter1998a,Kang2010a,Parsons2012a}, \formaldehyde
\citep{Ginsburg2015a,Ginsburg2016b}, and HI \citep{Koo1997a}.

Millimeter continuum surveys helped reinvigorate interest in this cloud.  While
in CO, the W51 cloud looks like many other regions in the Galactic plane, in
dust the main star formation site, W51A, stands out as particularly luminous,
comparable only to W49A and Sgr B2 \citep{Ginsburg2012a,Csengeri2013a}.  The
W51 IRS2 and e1/e2 regions are among a small handful ($<10$) of regions that
are capable of forming a $M>10^4$ \msun cluster in our Galaxy
\citep{Ginsburg2012a,Bressert2012a,Urquhart2014a}.

\subsection{Geography}
The W51 cloud appears peculiar in the overall Galactic position-velocity diagram
\citep{Dame2001a}.  Most of the cloud exists at `forbidden' velocities above
the tangent velocity, $v>v_{tan}$.  Such a high line-of-sight velocity means
the cloud complex is most likely within a few hundred parsecs of the tangent
point, but it also implies that either a cloud-cloud collision or a
gravitational interaction with a deep potential, e.g., a spiral arm, has
accelerated the cloud complex \citep{Ginsburg2015a}.  The parallax distance to
masers associated with W51A, the main star-forming component of W51, have been
measured, giving $D=5.41^{+0.31}_{-0.28}$ kpc to W51 e2 and e8 \citep{Sato2010a}
and $D=5.1^{+2.9}_{-1.4}$ kpc to W51 IRS2 \citep{Xu2009a}.  These distances
put W51 in the Carina-Sagittarius arm \citep{Reid2009a,Reid2014a}.

\Figure{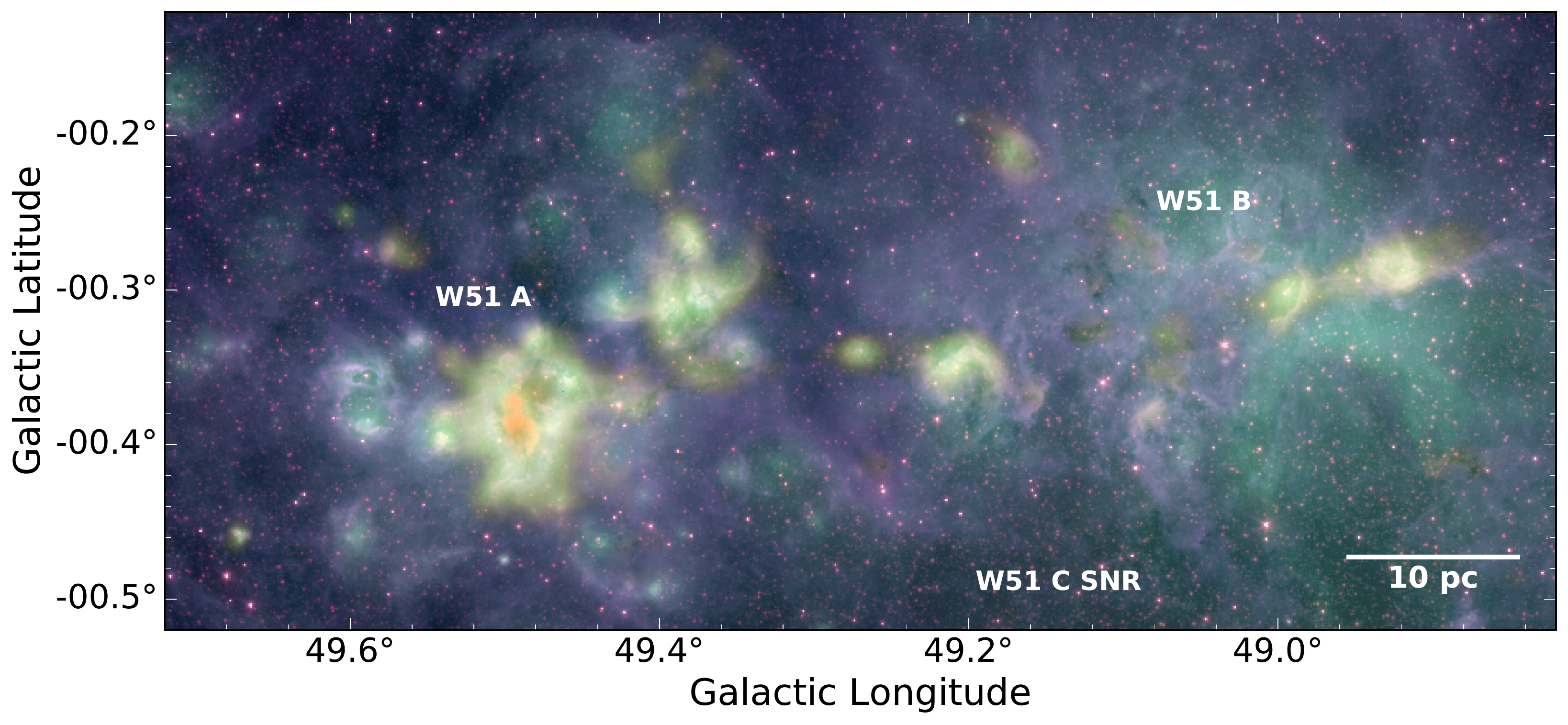}
{A color composite of  W51  with major regions, W51 A, B, and C,
labeled.  W51 A contains the protoclusters W51 Main and W51 IRS 2; these
are blended in the light orange region around 49.5-0.38.  The
blue, green, and red colors are WISE bands 1, 3, and 4 (3.4, 12,
and 22 \um) respectively.  The yellow-orange semitransparent layer is from the
Bolocam 1.1 mm Galactic Plane Survey data \citep{Aguirre2011a,Ginsburg2013a}.
 This figure was reproduced from \citet{Ginsburg2015a}.}
{fig:w51large}{1}{18cm}

The W51 cloud complex lies at a latitude $b\sim-0.3$, which given our vantage
point 25 pc above the Galactic plane means that W51 is very close to the
Galactic midplane.  The W51B cloud's elongated dust filament
\citep{Koo1999a,Wang2015a} is parallel to  the Galactic equator, making the
W51B filament a potentially more evolved analog of well-known filamentary
``spines'' like Nessie \citep{Goodman2014b}.

Despite its distance, but perhaps in part because of our vantage point, there
is very little molecular gas along the line of sight to W51.  Because of the
bright, compact \hii regions in the cloud, the limits on any such features
are fairly strict, $N(\hh)<10^{21}$ \persc \citep{Indriolo2012a}.  Most
of the molecular line emission, and absorption, is local to W51, i.e., it
is $\sim 5.5 \pm 1$ kpc in that general direction.

There is evidence that the higher velocity clouds, the $\sim55-65$ \kms W51 A /
W51 Main cloud and the 68 \kms cloud, are closer to each other than their
velocities imply, and they are interacting
\citep{Carpenter1998a,Bieging2010a,Ginsburg2015a}.  The lower-velocity clouds
around $\sim40$ \kms are  behind the other clouds, though it is unclear whether
they are  part of the W51 complex.  Their latitude and on-the-sky proximity to
the rest of W51 hints that they are related.


Along the molecular ridge that defines W51A and W51B, there are hints of
interaction with the supernova remnant W51C.  Direct evidence of this
interaction is observed within W51B through high-velocity CO and HI and from
shocked SiO emission
\citep{Koo1997b,Koo1997c,Aleksic2012a,Brogan2013a,Dumas2014a} and  OH masers
\citep{Brogan2013a}.  The extended W51 cloud is at least in part being
disrupted by this supernova remnant, as evidenced by a  lack of CO
emission toward much of the area filled by the radio remnant
\citep{Carpenter1998a,Bieging2010a,Parsons2012a}.

\Figure{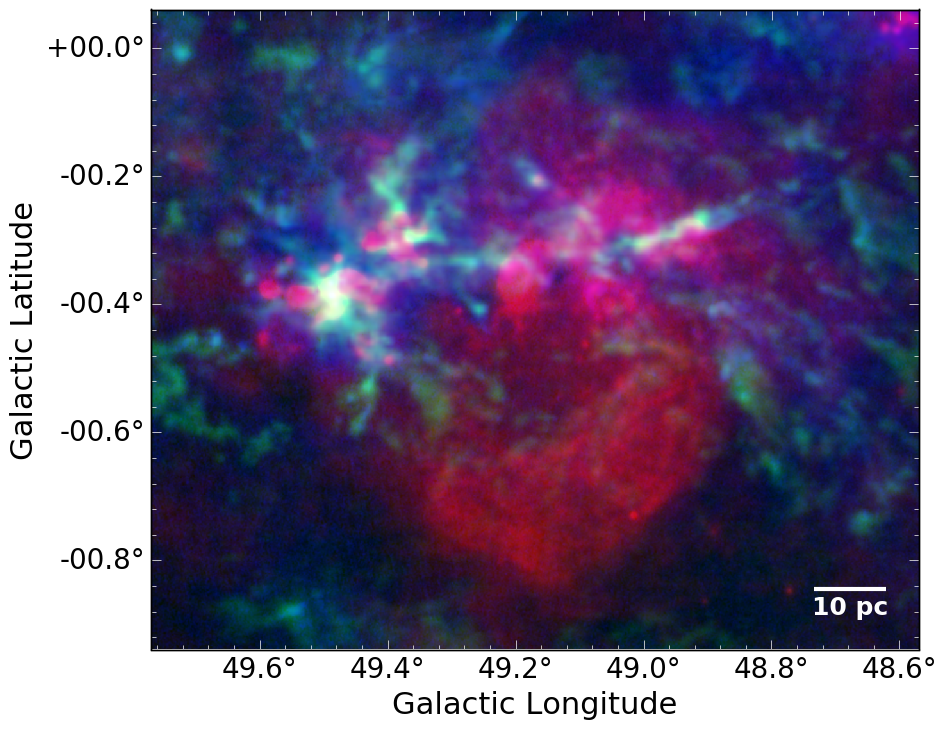}
{Another RGB figure of the W51 region on an even larger scale.
The red layer shows the 90 cm continuum measured with the VLA
\citet{Brogan2013a}, highlighting the W51 C supernova remnant as the extended
haze that dominates the image.  Blue shows ATLASGAL 870 \um continuum.  Green
shows the integrated $^{13}$CO emission from 30 to 90 \kms from the Galactic
Ring Survey \citep{Jackson2006a}.}
{fig:rgb2}{1}{18cm}

The total infrared luminosity of the W51 protocluster complex has been
estimated using IRAS and KAO, $L_{bol} \sim 9.3\times10^6 (D/5.4
\mathrm{kpc})^2 \lsun$ \citep{Harvey1986a,Sievers1991a}, though Herschel data
suggest the total IR luminosity might be a few times larger
\citep{Wang2015a,Ginsburg2016a}.

\subsection{\hii regions}
The W51 complex is best known as a collection of bright cm-wavelength radio
sources, which trace \hii regions.  \citet{Mehringer1994a} identified $\sim20$
independent \hii regions with the VLA.  Because these regions are so bright, it
has been possible to measure radio recombination line emission at high
significance, which allowed measurements of their electron temperature $T_e
\approx 7500$ K \citep{Mehringer1994a,Ginsburg2015a}.

\subsection{High mass star formation within W51A}
The W51A complex is the most actively studied part of the cloud complex, as it
contains some of the densest and most chemically complex gas in the Galaxy.
Because of their high millimeter brightness, the W51 IRS2 and W51 e1/e2 regions
have been the target of many millimeter studies revealing \methylcyanide
\citep{Remijan2004a,Remijan2004b}, \methylformate \citep{Demyk2008a}, H$_2$Cl+
\citep{Neufeld2015b}, HF \citep{Sonnentrucker2010a}, [NII]
\citep{Persson2014a}, and many other species.

Early radio studies of the region revealed the extremely bright and compact
source W51e2 and the similarly bright but more diffuse W51 IRS2 region
\citep{Mehringer1994a}.  Early follow-up of these regions showed signs of
infalling gas and ongoing accretion onto the \hii regions
\citep{Zhang1997a,Young1998a,Keto2008b}, though later higher-resolution
observations revealed that the infall is likely onto molecular cores adjacent
to the \hii regions \citep{Shi2010a,Shi2010b,Goddi2015a,Goddi2016a}.

The W51A complex is rich in molecular masers.  All three of the high-mass
protostars, W51e2, W51e8, and W51 North contain the usual OH
\citep{Etoka2012a}, H$_2$O \citep{Genzel1981a,Imai2002a,Eisner2002a}, and
\methanol masers \citep{Phillips2005a,Etoka2012a}.  However, W51 North also
contains some rare or unique masers: SiO \citep{Morita1992a,Eisner2002a} and
NH$_3$ \citep{Brown1991a,Gaume1993a,Henkel2013a}.  Both of these classes of
masers are only detected toward a few star forming regions in the Galaxy, with
5-8 known in SiO \citep{Morita1992a,Zapata2009c,Ginsburg2015b,Cordiner2016a},
and only 5 in \ammonia \citep{Madden1986a,Walsh2007a}.  W51 e2e and e8 also have
\ammonia (9,6) masers \citep{Pratap1991a}.  Recent high-resolution
ALMA data add to the mystery of these masers, revealing that
the \ammonia and SiO masers come from different sources within W51 North
separated by only 0.7\arcsec \citep[4000 AU; Goddi et al in prep;][]{Goddi2015a}.

\Figure{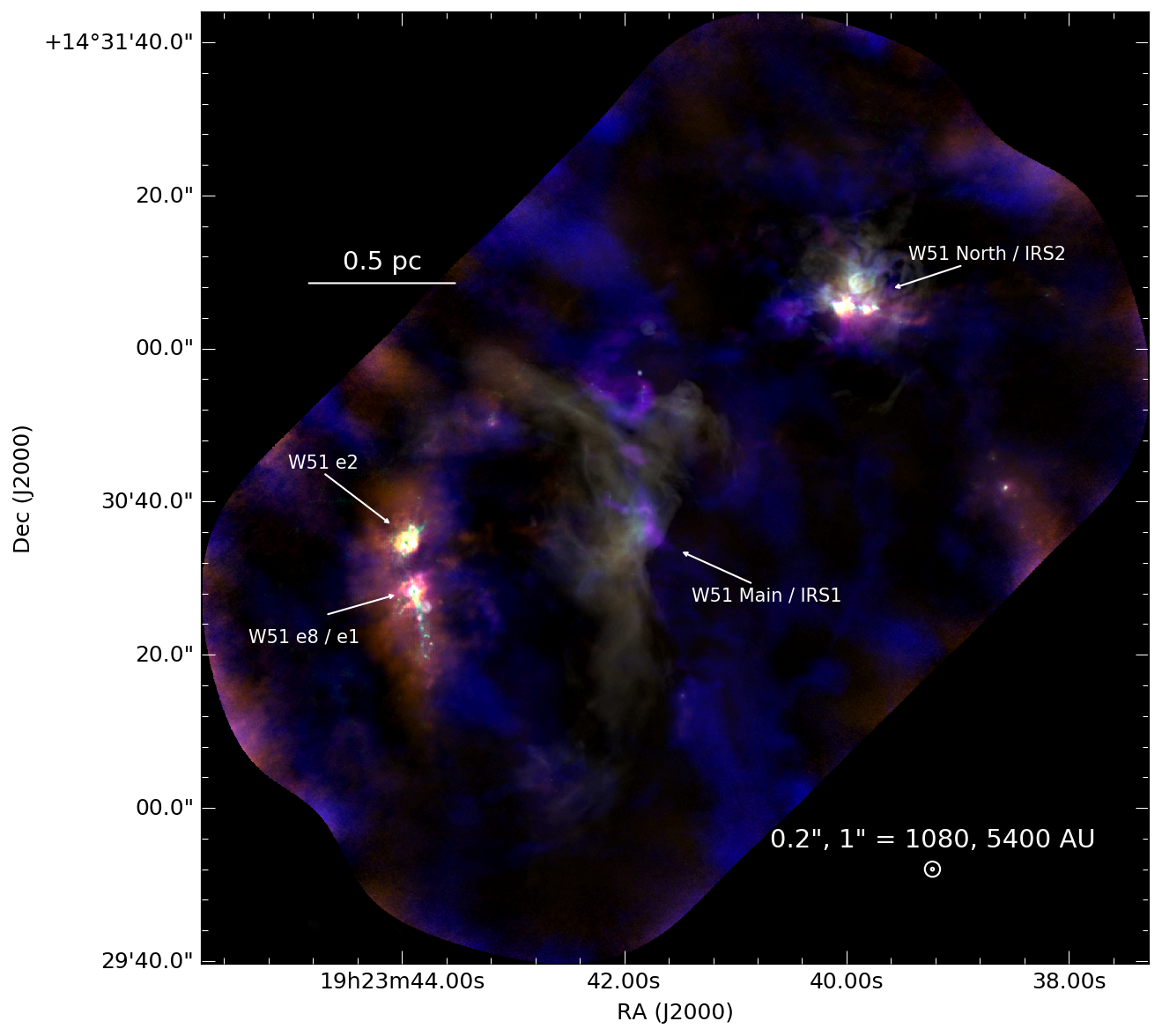}
{An overview of the W51A region as seen by ALMA and the VLA.  
The most prominent features are labeled.  W51 e8 is a mm dust source, while W51 e1
is the neighboring \hii region.  Similarly, W51 IRS2 is the \hii region, and
W51 North is the brightest mm source in that area.  The colors are a composite
of millimeter emission lines: CO in blue, \methanol in orange, and HC$_3$N in
purple \citep{Ginsburg2017a}.  The 1.3mm continuum is shown in green.  The
white hazy emission shows VLA Ku-band free-free continuum emission
\citep{Ginsburg2016b}.}
{fig:overview}{1}{18cm}

\subsection{The stellar population of W51A}
The W51A region has a luminosity corresponding to a 5000-10000 \msun
cluster.  Near-infrared observations have identified a small subset of this
population.  Early infrared studies found about 20 OB stars in the
gas-rich,
radio-bright region \citep{Okumura2000a,Kumar2004a}.  These early observations
were limited by confusion with the \hii regions, especially in W51 Main and W51
IRS2.  Those limitations were partly overcome with adaptive optics imaging,
which provided the first glimpse into the dense cluster formed within W51 IRS2
and revealed very luminous and young O3/O4 stars and at least one currently
accreting O-star \citep{Figueredo2008a,Barbosa2008a}.

Despite significant progress in identifying the stellar populations within W51,
the most deeply embedded - but nonetheless main-sequence - stellar groups may
not yet be identified \citep{Ginsburg2016a}.  X-ray observations reveal a
genuine cluster associated with W51 IRS2 \citep{Townsley2014a}, indicating the
presence of many stars that were not detected in the infrared.  Radio
observations show pointlike sources spread around the most luminous \hii
region \citep{Mehringer1994a,Ginsburg2016a}, W51 Main, but in this region
infrared point sources have been difficult to detect and there is no clear
X-ray cluster; it appears that this \hii region may be illuminated by a broader
OB association that has so far gone mostly uncharacterized
\citep{Ginsburg2016a}.

\subsection{The stellar population elsewhere}
A few independent studies of the stellar population throughout W51 have been
performed in the near-infrared \citep{Goldader1994a,Okumura2000a,Kumar2004a}.
There are some hints of a slightly top-heavy initial mass function in the W51
IRS2 region \citep{Okumura2000a}.  \citet{Kang2009a} studied the embedded
stellar population with Spitzer, finding that most of the more massive
protostars are associated with or very nearby \hii regions.  Besides the W51
IRS2 cluster (G49.5-0.4), \citet{Kumar2004a} found an additional three embedded
clusters in the lower-longitude W51 B region with ages 1-3 Myr and masses
2-10\ee{3} \msun.  The existence of these older clusters supports the
hypothesis that star formation has been continuously ongoing for $\sim10$ Myr,
and led those authors to suggest that star formation in W51 was triggered by
interaction with a spiral arm, a notion that is supported by gas kinematics
studies \citep{Ginsburg2015a}.

\subsection{Peculiar and notable objects within W51:\\ W51 Main}

\subsubsection{W51 e2}
The most famous source in W51, in large part due to its interesting
astrochemistry, is W51e2.  The source was originally noted as a hypercompact
\hii region \citep{Mehringer1994a}, as it is the brightest compact source in
the area.  Recent studies have revealed that it breaks into at least two
sources, e2w being the \hchii region and e2e a nearby hot molecular core
\citep{Shi2010a,Shi2010b,Goddi2016a,Ginsburg2017a}.  The hot core is one of the
major targets for discovery of complex chemical species, coming in just behind
Orion BN/KL and Sgr B2, likely because of this unique interaction between a
very bright radio centimeter continuum backlight and a hot molecular core.

The magnetic field direction has been measured with polarization toward e2
\citep{Tang2009a}, and it follows the outflow.  \citet{Etoka2012a} measure
a B-field strength of 2-7 mG with OH masers.

The \hchii region e2w is an appealing target for studies of compact ionized gas
since it is so bright.  \citet{Keto2008b} and \citet{Klaassen2009a} observed
possible rotation in the ionized gas, suggesting the presence of a disk.  More
recent observations \citep{Goddi2016a,Ginsburg2016a} have called this
interpretation into question, but it nonetheless represents an important target
for this type of study.

\subsubsection{W51 e8}
There is a cluster of compact \hii regions adjacent to W51e2
\citep{Mehringer1994a}.  The brightest is the extended W51e1 ultracompact \hii
region, but the most molecularly rich is the W51 e8 hot core.  Unique among the
hot molecular cores in W51, e8 has a radio continuum detection and therefore
contains at least some free-free or synchrotron emission.  The W51 e sources
lay along a common molecular ridge and are very likely to be within $<1$ pc of
each other \citep{Ginsburg2017a}.

\subsection{Peculiar and notable objects within W51:\\ W51 IRS 2}
W51 IRS2 is the northern of the two major protoclusters in W51.  It is powered
by at least one O3/O4 star \citep{Barbosa2008a} and contains a luminous compact
\hii region (Figure \ref{fig:irs2}).

\Figure{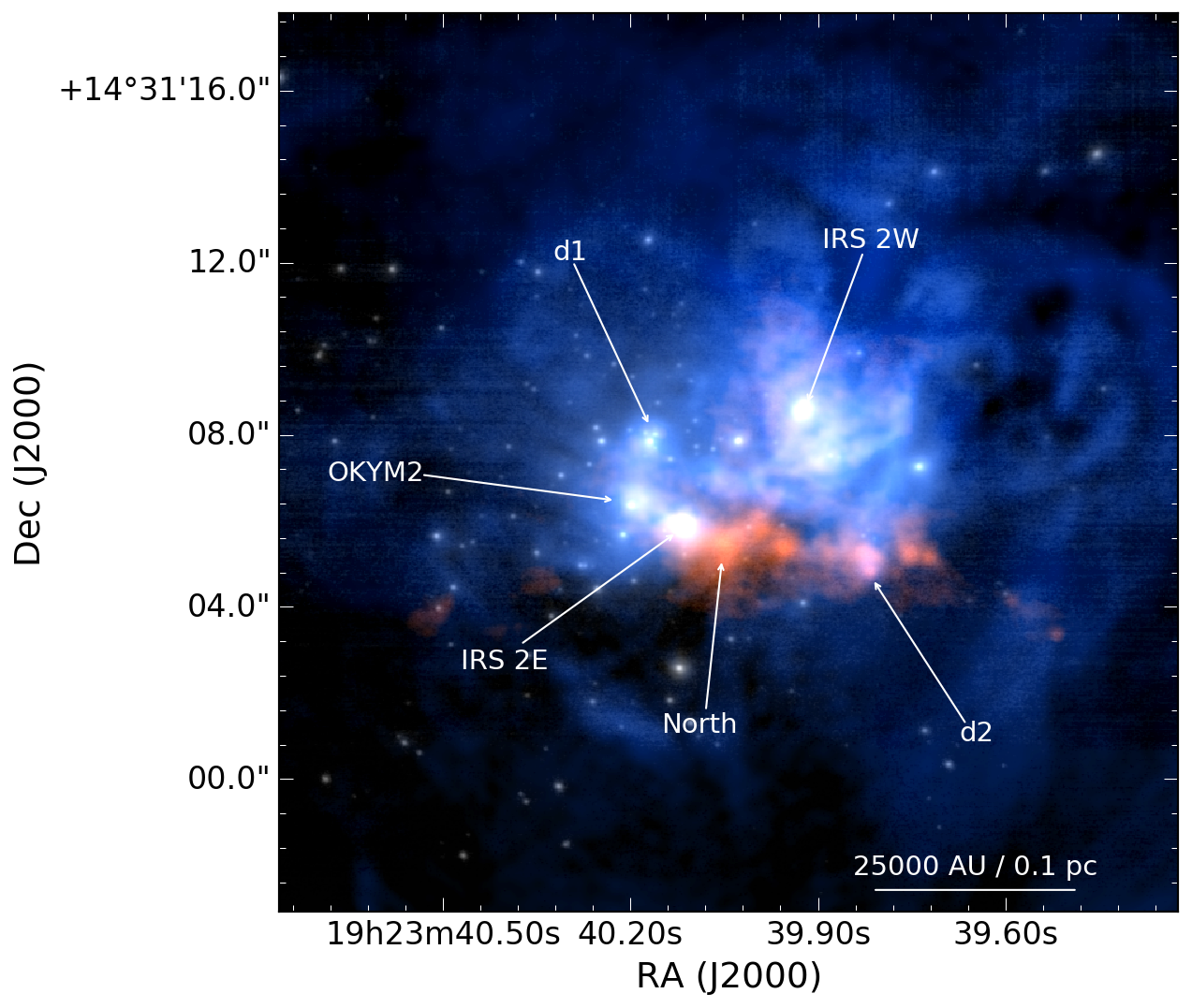}
{An image of the W51 IRS2 region, consisting of 2 cm VLA Ku-band continuum
\citep[blue;][]{Ginsburg2016a}, 1.3 mm ALMA Band 6 continuum
\citep[orange;][]{Ginsburg2017a}, and near-infrared VLT NAOS CONICA K-band
adaptive optics \citep[white;][]{Figueredo2008a} data.  The millimeter emission
traces a dust ridge containing W51 North (\S \ref{sec:north}) and d2 (\S
\ref{sec:d2}).
}{fig:irs2}{1}{16cm}

\subsubsection{W51 North}
\label{sec:north}
The W51 North hot core is directly adjacent to the W51 IRS2 \hii region and is
at least partly embedded within it.  This core contains one of the rare SiO
masers, so far detected toward only a handful ($\lesssim8$) high-mass
star-forming regions \citep{Eisner2002a,Zapata2009c,Ginsburg2015b}.
It is one of only a small number
\citep[$\lesssim4$][]{Walsh2007a,Hoffman2011b,Hoffman2011a} of known
metastable and non-metastable \ammonia maser sources
\citep{Madden1986a,Mauersberger1987a,Wilson1988a,Wilson1990a,Gaume1993a,Henkel2013a,Goddi2015a}.
It was recently detected at 25 \um, implying that it contains warm
dust \citep{Barbosa2016a}.

\subsubsection{W51d2}
\label{sec:d2}
W51d2 is the brightest hypercompact \hii region in the IRS2 area.  It is the
only \hchii region that is clearly associated with surrounding molecular gas
\citep{Zhang1997a,Ginsburg2017a}.  This source also contains one of the rare
non-metastable \ammonia masers (see citations for W51 North).  It is detected
at 25 \um, but not at shorter wavelengths \citep{Barbosa2016a}.

\subsubsection{W51 IRS2E}
IRS2E is a high-mass protostar with a detected circumstellar disk
\citep{Barbosa2008a,Figueredo2008a}.
X-ray emission, notably in the 6.5 keV iron fluorescence line, has been detected
toward this source and is variable \citep{Townsley2005a,Townsley2014a}.

\subsubsection{The Lacy Jet}
\label{sec:lacyjet}
There is a bipolar, highly symmetric outflow with Z-symmetry in the IRS2 area.
This outflow is unique in that it was first detected entirely in ionized gas
emission \citep{Lacy2007a}, with the molecular counterpart detected only with
ALMA observations a few years later \citep{Ginsburg2017a}.  Despite its
symmetry, which clearly pinpoints an origin position, no obvious centimeter or
millimeter line or continuum source has been detected at the base of the
outflow; there is a source very near the base, but offset by at least a few
hundred AU.

\Figure{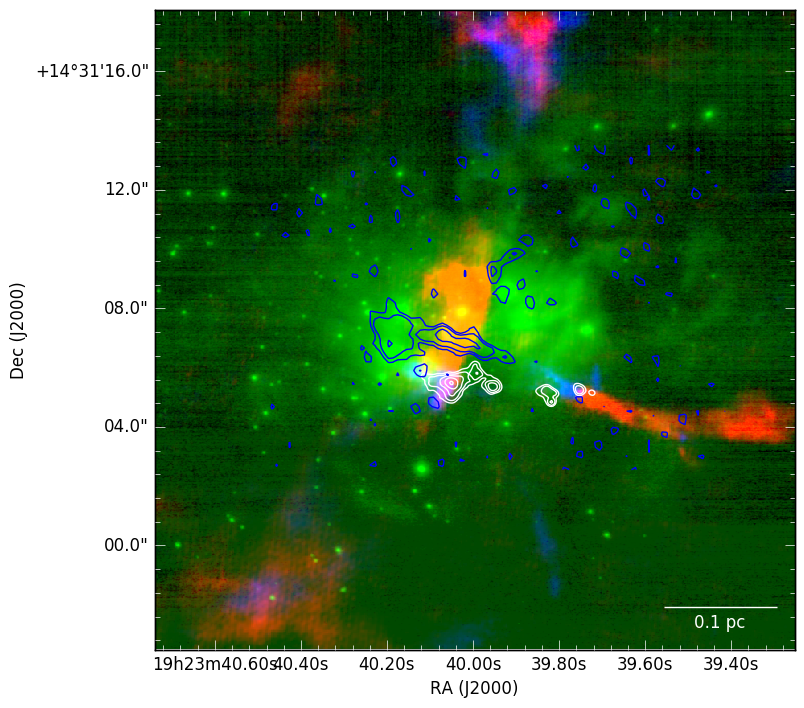}
{The W51 IRS2 region with a near-infrared K-band background in green,
\citep{Figueredo2008a}, CO outflows in red and blue, ALMA 1.4 mm  in
white contours, \citep{Ginsburg2017a} and high-velocity H77$\alpha$ in blue
contours \citep{Ginsburg2016a}.  The image shows two clear outflows, the
north-south large-scale flow from W51 North and the much more clearly bipolar
Lacy jet (\S \ref{sec:lacyjet}) that has no clear origin but exhibits an
S-shape symmetry when the ionized gas is included.  The symmetry is matched in
velocity.}
{fig:lacyjet}{1}{15cm}

\subsection{Peculiar and notable objects within W51: Others}
\subsubsection{OMN2000 LS1}
There is a P Cygni supergiant (an evolved high-mass star that may be a Luminous
Blue Variable, though no eruptive outbursts have been detected) in the W51
complex, whose presence indicates that high-mass star formation has been
ongoing for $>10$ Myr in the cloud \citep{Clark2009a}.  The lack of other
evolved massive stars of a similar age suggests that star formation has
been accelerating.

\subsubsection{The W51 C Supernova Remnant}
W51 C is an extended region of radio emission to the west of W51 A and mostly
south of W51 B (Figure \ref{fig:rgb2}).  It is a source of X-rays
\citep{Koo1995a,Koo2002a,Koo2005a,Hanabata2013a}, gamma rays \citep{Abdo2009a},
and high-energy cosmic rays \citep{Fang2010a}.  It is interacting with the W51
B cloud \citep{Koo1997a,Koo1997b,Brogan2013a,Ginsburg2015a}.  It has
evacuated much of the volume of the W51B region of molecular gas and heated
much of the remaining gas \citep{Parsons2012a,Ginsburg2015a}.  The interaction
region between the supernova remnant is visible in very high velocity emission
of molecular and atomic tracers (HI and CO) and exhibits OH maser emission
\citep{Brogan2013a}.

\section{Data on W51}
W51 has some additional appeal as a laboratory because there is plenty of data
available on it.  It is included in most Galactic plane surveys (UKIDSS, BGPS,
ATLASGAL, HiGAL, Spitzer GLIMPSE/MIPSGAL, GLOSTAR, CORNISH, THOR, BU-GRS) and
has many additional data sets that can be found online.  Much of the data shown
in this review is available on the Dataverse
(\url{https://dataverse.harvard.edu/dataverse/W51_ALMA}).

\section{Conclusion}
W51 is an excellent laboratory for the study of high-mass star formation and
high-mass cluster formation because of its unique location in the Galaxy.  The
complex is in one of the least crowded parts of the Galactic plane and it is
the closest that contains protoclusters with $M\geq10^4$ \msun.

\section{Acknowledgements}
I thank Miller Goss and John Bally for their helpful reviews.

\input{solobib}

\end{document}

%% file: preface.tex
\input{apjmacros.tex}

\usepackage{latexsym}		
\usepackage{graphicx}		
\usepackage{rotating}		
\usepackage{natbib}  
\usepackage{savesym}
\usepackage{amssymb}
\usepackage{morefloats}
\savesymbol{doublespace}
\usepackage{xspace}
\usepackage{color}
\usepackage{mdframed}
\usepackage{url}
\usepackage{subfigure}
\usepackage{grffile}
\usepackage{standalone}
\standalonetrue
\usepackage{import}
\usepackage[utf8]{inputenc}
\usepackage{longtable}
\usepackage{booktabs}
\usepackage[yyyymmdd,hhmmss]{datetime}
\usepackage{fancyhdr}
\usepackage[colorlinks=true,citecolor=blue,linkcolor=cyan]{hyperref}
\usepackage{ifpdf}

\input{macros}		

%% file: apjmacros.tex
\newcommand{\titlerunning}[1]{\shorttitle{#1}}
\newcommand{\authorrunning}[1]{\shortauthors{#1}}

\newcommand*\inst[1]{\unskip\hbox{\@textsuperscript{\normalfont$#1$}}}

\newcommand*\institute[1]{
  \begingroup
    \let\and\relax
    \renewcommand*\inst[1]{}%
    \renewcommand*\thanks[1]{}%
    \renewcommand*\email[1]{}%
  \endgroup
  \newcommand{\institutions}{#1}
}%

\renewcommand{\ion}[2]{\textup{#1\,\textsc{\lowercase{#2}}}}

%% file: macros.tex
\newcommand{\msun}{\ensuremath{M_{\odot}}\xspace}			

\newcommand{\lsun}{\ensuremath{L_{\odot}}\xspace}			
\newcommand{\hh}{\ensuremath{\textrm{H}_{2}}\xspace}			

\newcommand{\formaldehyde}{\ensuremath{\textrm{H}_2\textrm{CO}}\xspace}

\newcommand{\methanol}{\ensuremath{\textrm{CH}_3\textrm{OH}}\xspace}

\newcommand{\methylcyanide}{\ensuremath{\textrm{CH}_{3}\textrm{CN}}\xspace}

\newcommand{\methylformate}{\ensuremath{\textrm{CH}_{3}\textrm{OCHO}}\xspace}

\newcommand{\hchii}{\ion{HCH}{ii}\xspace}

\newcommand{\hii}{\ion{H}{ii}\xspace}

\newcommand{\kms}{\textrm{km~s}\ensuremath{^{-1}}\xspace}	

\newcommand{\persc}{\ensuremath{\textrm{cm}^{-2}}\xspace}

\newcommand{\um}{\ensuremath{\mu \textrm{m}}\xspace}    

\newcommand{\ammonia}{NH\ensuremath{_3}\xspace}

\def\ee#1{\ensuremath{\times10^{#1}}}

\def\eqref#1{Equation \ref{#1}}

\def\Figure#1#2#3#4#5{
\begin{figure*}[!htp]
\includegraphics[scale=#4,width=#5]{#1}
\caption{#2}
\label{#3}
\end{figure*}
}

\newenvironment{rotatepage}%
{}{}

%% file: authors.tex
\newcommand{\nrao}{$^{1}$}

\author{
Adam Ginsburg{\nrao},
\begin{flushleft}
\institutions
\end{flushleft}
        }

\institute{
    {\nrao}{\it{National Radio Astronomy Observatory, Socorro, NM 87801 USA\\
                      \email{aginsbur@nrao.edu} 
                      }} \\
    }

%% file: solobib.tex
\ifstandalone
\bibliographystyle{apj_w_etal}  
\bibliography{bibdesk}      
\fi